\documentclass[aps,prb,twocolumn]{revtex4}
\usepackage{graphicx}

\begin{document}

\title{Lattice Dynamics in PbMg$_{1/3}$Nb$_{2/3}$O$_3$}

 \author{S. A. Prosandeev$^*$, E. Cockayne, B. P. Burton}
 \affiliation{Ceramics Division, Materials Science and
Engineering Laboratory, National Institute of Standards and
Technology, Gaithersburg, Maryland 20899-8520\\ $^*$ also Physics
Department, Rostov State University, 5 Zorge St., 344090 Rostov on
Don, Russia}

 \author{ S. Kamba and J. Petzelt}\affiliation{Institute of Physics,
Academy of Sciences of the Czech Republic, Na Slovance 2, 18221
Prague 8, Czech Republic}

\author{Yu.~Yuzyuk$^*$~ and R. S. Katiyar} \affiliation{University of Puerto
Rico, Dept. of Physics, P.O. BOX 23343, San Juan PR. 00931-3343
USA}

\author{S. B. Vakhrushev}
\affiliation{A. F. Ioffe Physical-Technical Institute,194021 St.
Petersburg, Russia}

\date{\today}

\begin{abstract}
Lattice dynamics for five ordered
$\mathrm{PbMg_{1/3}Nb_{2/3}O_3}$~ supercells were calculated from
first principles by the frozen phonon method. Maximal symmetries
of all supercells are reduced by structural instabilities.
Lattice modes corresponding
to these instabilities, equilibrium ionic positions, and infrared
reflectivity spectra were computed for all supercells. Results are
compared with our experimental data for a chemically disordered
PMN single crystal.
\end{abstract}

\maketitle

\section{Introduction}
Lead magnesium niobate PbMg$_{1/3}$Nb$_{2/3}$O$_3$~(PMN)  is an
ABO$_3$-perovskite that exhibits relaxor ferroelectric (RFE)
properties, such as a broad, frequency sensitive, dielectric
permittivity peak at
$T_{max}$ near room temperature.\cite{Smolenski,Cross} It was
originally suggested that RFE properties are associated with a
``diffuse phase transition" that reflects a \emph{random} spatial
distribution of Curie temperatures, \cite{Smolenski} originating
from chemical inhomogeneities; however, the implied chemical
segregation of  Mg$^{2+}$~ and Nb$^{5+}$~ ions on perovskite
B-sites has not been observed.  Rather, transmission electron
microscopy (TEM) indicates chemical short-range order (SRO) such
that 1:1 ordered domains, 20-50 \AA~across, are embedded in a
disordered matrix. \cite{Krause}
Thus, the microstructure is characterized by
fluctuations of a 1:1 chemical order parameter on the
20-50~\AA~length scale, and not by compositional fluctuations about
Mg:Nb=1:2, as previously suggested.\cite{Husson1,Husson}

The ``random site model (RSM)" for the ordered domains,
\cite{Galasso,Krause1,Akbas,Davis00JPCS,Davis00Aspen} has
NaCl-type Mg/Nb ordering with one B-site occupied by Nb$^{5+}$,
and the other by a random mixture of 1/3~Nb$^{5+}$+2/3~Mg$^{2+}$.
Note that the phrase ``random-site model" is only used for an
idealized model. The phrase ``1:1 phase" is used for real crystals
that presumably have some SRO chemical correlations which are
deviations from $randomness$~ on the Mg$^{2+}$-rich sites.

Burns and Dacol  \cite{Burns} observed that the refractive index
of PMN departs significantly from a linear temperature dependence
below a characteristic temperature, $T_{B} \sim 630K$~ (Burns
temperature).  They interpreted these data as indicating that
polar nanoclusters (PNC) condense at $T < T_{B}$; and $T_{B}$~ is
now generally accepted as the crossover between
RFE and paraelectric (PE) states.
Recent TEM studies \cite{Jin, Miao} have attempted to correlate
chemical SRO with PNC.

Chemical SRO makes PMN inhomogeneous at a length-scale that
affects vibrational mode activities. Mode activities in
inhomogeneous media are determined by the range of interatomic
forces, typically a few lattice parameters, and they do not depend
on the wavelength of the probing radiation. However, the probing
radiation wavelength does determine the volume in which the
spectroscopic response is averaged. If, as in PMN, the
inhomogeneities are smaller than the wavelength, an effective
medium approach is justified.

Room-temperature IR reflectivity spectra of PMN single crystals
were first published in the 1970s. \cite{Burns,Karamyan} Karamyan
reported three polar modes, characteristic of cubic perovskites.
Burns and Dacol, however, reported many more modes, which they
explained as ``two-mode behavior".\cite{Barker} Subsequent studies
of PMN ceramics\cite{Reaney} agreed with Karamyan's spectrum, but
the resolution was poor.

The interpretation of Raman spectra from PMN is also controversial
(e.g. \cite{Siny00,Toulouse}).  The Raman spectra of PMN and PST
are similar, \cite{Idink,Siny00} and it was assumed that the main
features of both reflect $Fm\overline{3}m$~ symmetry (locally for
PMN) as in the 1:1 phase. In $Fm\overline{3}m$, the Raman
active modes are $A_{1g}$, $E_g$, and 2$F_{2g}$: $A_{1g}$~ and
$E_g$~ modes are observable in parallel (VV) geometry; 2$F_{2g}$~
modes are observable in crossed-polarized (VH) geometry. Polarized
Raman studies and Raman spectral intensities support TEM results
in suggesting that PMN has 1:1 chemical SRO in a disordered
matrix.

The objectives of this study are: (1) to perform a comparative
supercell stability analyses in PMN, including symmetry-breaking
relaxations [{\it i.e.} finding ground states (GS) with symmetries
that are lower than those dictated by chemical ordering]. (2) to
compute the lattice dynamics for the same set of ordered PMN
supercells, and to compare the results with experimental data,
{\it e.g.} by comparing simulated IR reflectivity spectra with the
experimental one.  The goal is to find which ordered supercell
most closely approximates the experimental case of local 1:1
order.

\section{Infrared  spectra}\label{IR}

IR and THz measurements were performed on a PMN single crystal,
down to liquid helium temperatures, where all polar modes are
well distinguished.  A disk of 9~mm diameter and 3~mm thickness
was cut from a PMN single crystal, and polished for specular
IR reflectivity measurements.  A time-domain THz spectrometer
was used to determine the complex dielectric response
$\varepsilon^{*}(\nu)$ in the submillimeter and near-millimeter
ranges. This spectrometer uses femtosecond laser pulses to
generate THz radiation via optical rectification on a ZnTe single
crystal, with an electro-optic sampling detection technique.
Low-temperature spectra were taken in transmission configuration
from 100-900 GHz, where the thin (100 $\mu$m thick) plane-parallel
plate was semi-transparent. Room temperature dielectric response
up to 2.5 THz was calculated from the THz reflectance data.
\cite{Bovtun03} The unpolarized near-normal reflectivity spectra
were taken with a Fourier transform IR (FTIR) spectrometer Bruker
IFS 113v at $20 \leq T \leq 300~ K$~ in the spectral range 20-650
cm$^{-1}$; room-temperature spectra were measured up to 4000
cm$^{-1}$.

IR reflectivity spectra were fitted together with complex
dielectric THz spectra using a generalized-oscillator model with
the factorized form of the complex dielectric function:

\begin{equation}
\label{eps4p}
 \varepsilon^*(\nu) =
\varepsilon'(\nu)-\textrm{i} \varepsilon''(\nu)
 = \varepsilon_{\infty} \prod_{j=1}^{n}
\frac{\nu_{LOj}^{2} - \nu^2+\textrm{i}\nu\gamma_{LOj}}
{\nu_{TOj}^{2} - \nu^2+\textrm{i}\nu\gamma_{TOj}}
\end{equation}
where the dielectric function is related to reflectivity
R($\nu$) by
\begin{equation}
\label{Refl}
 R(\nu) =
\left|{\frac{\sqrt{\varepsilon^*(\nu)}-1}{\sqrt{\varepsilon^*(\nu)}+1}}\right|^2
\end{equation}
$\nu_{TOj}$ and $\nu_{LOj}$ are the transverse optic and
longitudinal optic (LO) frequency of the \textit{j}-th mode,
respectively; $\gamma_{TOj}$ and $\gamma_{LOj}$ the corresponding
damping constants. The high-frequency (electronic) permittivity
$\varepsilon_{\infty}$ was obtained from the frequency-independent
300K reflectivity above the phonon frequencies. The temperature
dependence of $\varepsilon_{\infty}$ is usually very small and was
neglected in our fits. The TO1 mode was fit to the more accurate
FTIR reflectivity. Below 30~cm$^{-1}$, the THz data are more
accurate than the FTIR data, so it was given greater weight in the
fit. The results are shown in Table~\ref{IRmodes}.  In particular,
the TO1 phonon frequency shifts from 88~cm$^{-1}$~ at 20~K down to
54 cm$^{-1}$~ at 300 K. Therefore we can call it soft mode (SM).

\begin{table*}[!htbp]
\caption{Parameters of the polar phonon modes in PMN, from
fits of IR and THz spectra at 20 and 300 K. Mode frequencies
$\nu_{TOi}$, $\nu_{LOi}$ and dampings $\gamma_{TOi}$,
$\gamma_{LOi}$ are in \ensuremath{\mbox{cm}^{-1}},
$\Delta \varepsilon_{i}$ is dimensionless,
$\varepsilon_{\infty}$=5.83. }
\begin{tabular}{|r | c c c c c||c c c c c |}\hline
  &\multicolumn{5}{c||}{20 K}&\multicolumn{5}{c|}{300 K}\\
  \hline
  No&$\hspace{0.2cm} \nu_{TOi} \hspace{0.2cm}$&\hspace{0.2cm}
  $\gamma_{TOi}$ \hspace{0.2cm}&\hspace{0.2cm} $\nu_{LOi}$ \hspace{0.2cm}&\hspace{0.2cm}
  $\gamma_{LOi}$ \hspace{0.2cm}&\hspace{0.2cm} $\Delta
  \varepsilon_{i}$ \hspace{0.2cm}&
  $\hspace{0.2cm} \nu_{TOi} \hspace{0.2cm}$&\hspace{0.2cm}
  $\gamma_{TOi}$ \hspace{0.2cm}&\hspace{0.2cm} $\nu_{LOi}$ \hspace{0.2cm}&\hspace{0.2cm}
  $\gamma_{LOi}$ \hspace{0.2cm}&\hspace{0.2cm} $\Delta
  \varepsilon_{i}$ \hspace{0.2cm}
 \\ \hline
 CM&36.2&44.8&62.5&56.2&129.2&24.2&69.2& & &539.1\\
 1 & 65.0&15.2&67.1&18.2&0.6&&&&&\\
(SM)2&88.2&22.0&122.9&32.7&15.7&54.3&38.3&102.0&17.3&79.6\\
 3&245.5&76.9&264.1&49.0&11.8&227.7&85.1&282.6&65.0&17.7\\
 4&272.3&59.4&382.1&60.9&4.1&287.7&55.0&334.2&45.0&0.9\\
 5&385.1&41.2&406.3&21.4&0.1&336.2&47.5&397.7&7.5&0.13\\
 6&434.0&93.7&455.3&22.0&0.3&427.9&55.7&442.3&35.4&0.2\\
 7&547.9&80.6&703.0&22.9&1.8&545.4&105.1&703.0&62.5&2.0\\
  \hline
  \hline
\end{tabular}
\label{IRmodes}
\end{table*}

The low-frequency part [so called central mode (CM)] of the 300K
spectrum, below polar phonon frequencies, was fit to an overdamped
three parameter oscillator model. This fit does not explain the
experimental low-frequency permittivity which is one order of
magnitude higher, owing to broad dispersion between the audio- and
microwave-frequency ranges.\cite{Bovtun03,Bovtun97,Levstik98} It
is well known that the distribution of relaxation frequencies in
this dispersion reaches up to the submillimeter range at room
temperature.\cite{Bovtun03,Levstik98} At 20~K, a new heavily
damped excitation appears near 30 cm$^{-1}$~
(Table~\ref{IRmodes}).

\begin{figure}[!htbp]
\resizebox{0.45\textwidth}{!} {\includegraphics{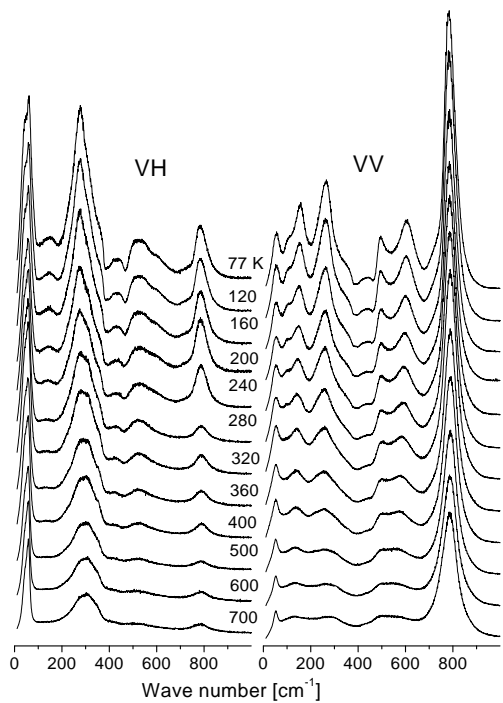}}
\caption{Temperature dependence of Raman intensity (corrected for
the temperature factor).} \label{figura}
\end{figure}

\section{Raman spectra}

Raman spectra were excited with polarized light from a coherent
INNOVA 99 Ar$^{ + }$ laser ($\lambda $ = 514.5 nm) and analyzed
with a Jobin Yvon T64000 spectrometer, that was equipped with a
charge coupled device (CCD). Polarized Raman spectra were measured
in backscattering geometry on the (1$\times $2$\times $3
mm$^{3})$~ sample.

Incident light was focused to a spot size of $\sim 3 \mu $m in
diameter, as measured by optical microscopy. Low- and
high-temperature micro-Raman measurements were made with a Linkam
FDCS 196 cryostat and Linkam TS1500 hot stage, respectively.
Measured Raman spectra were corrected for the Bose-Einstein
temperature factor.

Fig. \ref{figura} shows the temperature dependence of parallel
(VV) and crossed (VH) polarized Raman spectra in the temperature
interval 77-700 K. The main feature is that the frequencies of all
observed peaks are only weakly temperature-dependent. Below 270 K,
a partial depolarization occurs, due to the leakage of intense
peaks from the VV spectrum to VH.

All peaks are best resolved at low temperature, though most of
them can be traced up to 1000 K. As shown in Fig. \ref{figura} the
VV spectrum exhibits: a band at 54 cm$^{ - 1}$; an envelope of
strongly overlapping bands between 100 and 350 cm$^{ -1}$; a very
weak broad band at about 430 cm$^{ - 1}$, that is clearly seen
only at low temperature; strongly overlapping bands in the 500-600
cm$^{ - 1}$ range; and a very intense band at 780 cm$^{ - 1}$. In
the VH spectrum the low-frequency band is a doublet (45 and 62
cm$^{ - 1})$~ and intermediate bands in the 100-350 range cm$^{ -
1}$ are also strongly overlapping. The intensity of the 780 cm$^{
- 1}$ band increases abruptly at about 270 K, just below
$T_{max}$.  The 780 cm$^{ - 1}$ band was interpreted as a fully
symmetrical ($A_{1g}$) stretching vibration of oxygen octahedra
that originates from 1:1 chemically ordered regions. \cite{A1g}
The 500-600 cm$^{-1}$~ wide band, usually associated with $E_g$~
Raman active vibration, increases splitting when T decreases
\cite{Toulouse} below 350 K, and splits \cite{Kamba03} in I4mm
symmetry but not in R3m. Actually, any monoclinic distortions
split this mode either. Perfect 1:1 chemical ordering implies four
Raman-active modes: $A_{1g} + E_{g} + 2F_{2g}$. The measured Raman
spectra are much more complicated, however, which suggests that
both chemical SRO, and local symmetry reductions from lattice
instabilities, activate Raman modes that are inactive under
$Fm\overline{3}m$~ selection rules.

\section{Computational Methods}

Density functional theory (DFT) calculations for PMN were done
with the Vienna {\it ab initio} simulation package (VASP).
\cite{Kresse1,Kresse2} A plane wave basis set was used for
electronic wavefunctions and ultrasoft pseudopotentials were
employed.\cite{vanderb} Exchange and correlation energies were
calculated with the local density approximation (LDA).
Lattice-dynamical force constants were calculated via the frozen
phonon method. Berry's phase analyses were used to compute
dynamical charges, as implemented in VASP by M. Marsman.

\subsection{Ordered Supercells}

\begin{figure}
\resizebox{0.5\textwidth}{!}{\includegraphics{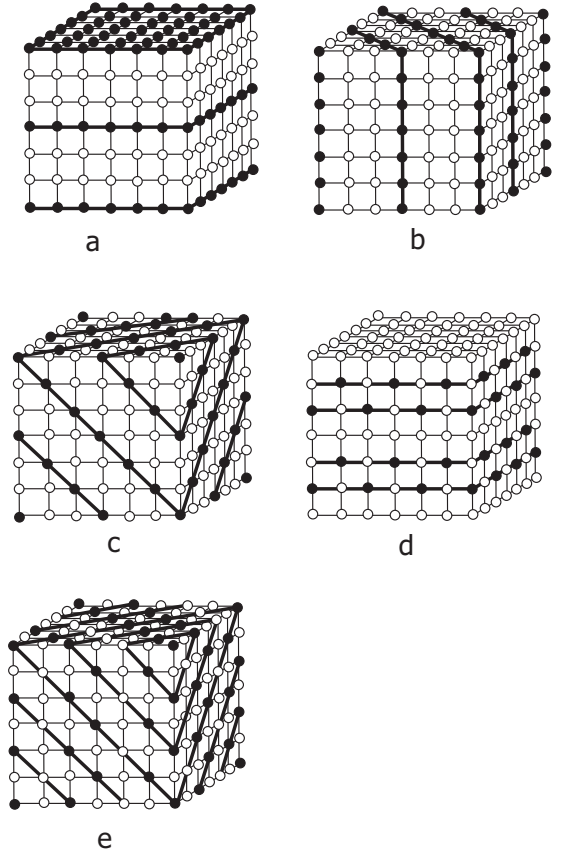}}
\caption{PMN supercells: [001]$_{NNM}$~(a), [110]$_{NNM}$~(b),
[111]$_{NNM}$~(c), [001]$_{NCC'}$~(d), [111]$_{NT}$~ (e)}
\label{structures}
\end{figure}

\begin{figure}
\resizebox{0.4\textwidth}{!}{\includegraphics{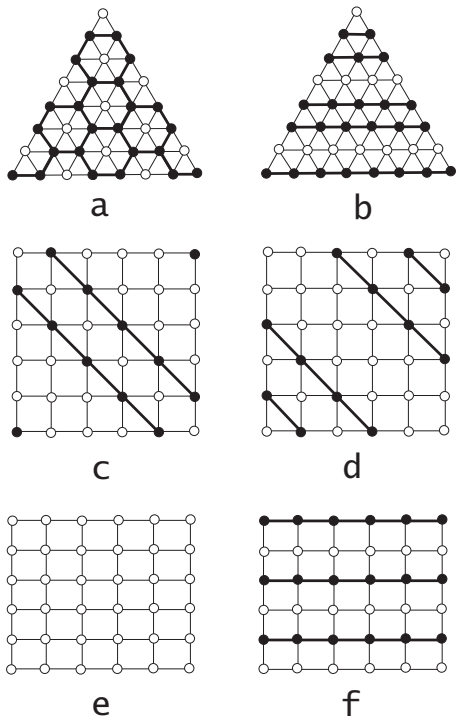}}
\caption{Filling the Mg$_{2/3}$Nb$_{1/3}$ layers in the
[111]$_{NT}$~ structure with Mg (black circles) and Nb (open
circles) in the direction of (a-b) [111], (c-d) [001], and (e-f)
[110].} \label{triangles} \label{triangles}
\end{figure}

\begin{table*}
\caption{Supercell space group symmetries $G$, formation energies
$\Delta E$ (kJ/mole, where mole is the Avogadro's number of a primitive
ABO$_3$~ unit cell, and relaxed cell parameters.)}
\begin{ruledtabular}
\begin{tabular}{c|cc|cccccc|cc|cccccc}
$System$ & $G_{Chem}$ & $\Delta E_{Chem}$ & $a$ & $b$ & $c$ &
$\alpha$ & $\beta$ & $\gamma$ & $G_{SSS}$ & $\Delta E_{SSS}$
&$a$&$b$&$c$&$\alpha$&$\beta$&$\gamma$
\\ \hline
     $[001]_{NNM}$ & $P4/mmm$        & 16.16&3.977 & 3.977 & 12.277 & 90 & 90 & 90
  &$Pm$&12.27&  4.050 & 3.965 & 12.250 & 90 & 90 & 89.84
 \\ $[110]_{NNM}$ & $Amm2  $         &14.33& 5.664 & 17.087 & 4.008 & 90 & 90 & 90
  &$Cm$&5.50 & 5.660  & 16.998 & 4.092 & 90 & 90.68 & 90
 \\ $[111]_{NNM}$ & $P\overline{3}m1$&14.11& 5.650 &  5.650 & 6.941 & 90 & 90 &120
  &$P1$&5.33 & 5.682 & 5.759 & 6.964 & 89.97 & 90.48 & 120.43
 \\ $[001]_{NCC'}$& $P4/nmm$         &11.51& 5.646 &  5.646 &12.044  &90 & 90 & 90
 & $P1$&2.99 & 5.645 & 5.665 & 12.197 & 90.15 & 89.93 & 89.85
 \\ $[111]_{NT}$ & $Immm$            &7.66& 5.615 &17.050  & 8.022  & 90 & 90 & 90
  &$P1$&0.00& (5.613 & 16.995 & 8.085 & 89.99 & 89.85 & 90.00)*
\end{tabular}
\end{ruledtabular}
*Not the primitive P1 cell; 60-atom cell allows comparison with Immm lattice
parameters.
\label{symmetry}
\end{table*}

\begin{table}[!htbp]\begin{ruledtabular}
 \caption{Ionic coordinates (in $\AA$) in 15 ion ordered supercells of PMN}
\begin{tabular}{c|ccc|ccc|ccc}
&
\multicolumn{3}{c|}{$[001]_{NNM}$}&\multicolumn{3}{c|}{$[110]_{NNM}$}
&\multicolumn{3}{c}{$[111]_{NNM}$}\\ \hline type& $x$  & $y$ & $z$
& $x$  & $y$ & $z$ & $x$  & $y$ & $z$
\\ \hline
 Mg     &  0.00 &-0.03 & 0.00 & 3.97 & 3.97 & 0.01 &-0.12 &-0.04 & 0.03 \\
 Nb$_1$ &  0.06 &-0.03 & 4.03 &-0.01 & 4.04 &-0.10 &-0.05 &-0.10 & 3.99 \\
 Nb$_2$ &  0.06 &-0.03 & 8.22 & 4.04 &-0.01 &-0.10 &-0.16 & 4.09 & 4.07 \\
 O$_1$  & -0.14 &-0.10 & 2.16 & 1.92 & 4.16 &-3.98 &-0.26 & 1.86 & 4.04 \\
 O$_2$  & -0.23 &-0.09 & 6.12 & 0.07 & 2.07 &-3.94 &-0.30 &-0.20 & 6.02 \\
 O$_3$  & -0.14 &-0.10 &10.09 & 6.00 &-0.16 &-3.92 & 1.79 & 3.88 & 3.99 \\
 O$_4$  &  1.99 &-0.12 & 0.00 &-0.16 & 6.00 &-3.92 &-0.29 &-0.18 & 2.06 \\
 O$_5$  & -0.14 & 1.93 & 0.00 & 4.16 & 1.92 &-3.98 &-0.37 &-2.08 & 4.01 \\
 O$_6$  &  1.94 & 0.01 & 4.18 & 2.07 & 0.07 &-3.94 & 1.83 &-0.17 & 3.98 \\
 O$_7$  &-0.17 & 1.91 & 4.23 & 0.11 & 0.11 &-2.02 & -0.23 & 3.84 & 5.99 \\
 O$_8$  & 1.94 & 0.01 & 8.07 &  0.01 & 4.01 &-1.99 &-0.23 & 5.96 & 4.07 \\
 O$_9$  &-0.17 & 1.91 & 8.02 &  4.01 & 0.01 &-1.99 &-2.14 & 3.76 & 4.04 \\
 Pb$_1$ & 2.17 & 2.04 & 1.60 & 1.77 & 1.77 &-2.27 &  2.17 & 2.13 & 2.15 \\
 Pb$_2$ & 2.26 & 1.98 & 6.12 & 6.24 &-2.33 &-2.23 &-1.97 & 2.14 & 2.04 \\
 Pb$_3$ & 2.17 & 2.04 &10.65 &-2.33 & 6.24 &-2.23 &-2.01 & 2.22 & 5.92
 \end{tabular} \label{poscar} \end{ruledtabular} \end{table}

\begin{table}[!htbp]
\begin{ruledtabular}
 \caption{Basic vectors for the computed structures}
\begin{tabular}{c|c|ccc}
   structure & multiplier & x & y & z \\ \hline
[001]$_{NNM}$ & 4
 & 1.012460 & 0.001455 & 0.000000\\
 && 0.001417 & 0.991364 & 0.000000\\
 && 0.000000 & 0.000000 & 3.062478\\
  \hline
 [011]$_{NNM}$  & 4
 & -1.002116 & 2.002737 & 0.004304\\
 &&  2.002737 &-1.002116 & 0.004304\\
 && -0.004184 &-0.004184 &-1.022970\\
 \hline
 [111]$_{NNM}$ & 5
 &-0.004329 &-0.809859 & 0.797156\\
 && 0.814385 & 0.814646 & 0.000288\\
 &&-0.806681 & 0.807082 & 0.798775\\
  \hline
[001]$_{NCC'}$ & 4
 & 0.999359 & -0.996947 & 0.000397\\
 && 1.002589 &  1.000282 &-0.002638\\
 && 0.000558 & -0.003683 & 3.049203\\
  \hline
  [111]$_{NT}$ & 4
 & 1.997368 &-1.006931 &-1.009787\\
 &&-1.006824 & 1.997432 &-1.009407\\
 && 0.992322 & 0.992308 & 0.001922
\end {tabular}
\end{ruledtabular}
\end{table}

\begin{table*}[!htbp]
\begin{ruledtabular}
 \caption{The dynamical charges of ions in the 15-ion supercells of PMN}
\begin{tabular}{c|ccc|ccccc|ccccccccc}
&\multicolumn{3}{c|}{$[001]_{NNM}$}&\multicolumn{5}{c|}{$[110]_{NNM}$}&\multicolumn{9}{c}{$[111]_{NNM}$}\\
\hline
 ion $i$ & $Z^*_{izz}$ & $Z^*_{ixx}$ &
$Z^*_{iyy}$&
$Z^*_{izz}$&$Z^*_{ixx}$&$Z^*_{ixy}$&$Z^*_{iyx}$&$Z^*_{iyy}$&$Z^*_{izz}$&
$Z^*_{izx}$&$Z^*_{izy}$&$Z^*_{ixz}$&$Z^*_{ixx}$&$Z^*_{ixy}$&$Z^*_{iyz}$&$Z^*_{iyx}$&$Z^*_{iyy}$
\\ \hline
 Mg     &  2.56  &  1.90  &  1.90 &  1.94 & 2.60 &-0.09&-0.09&
 2.60&2.72&-0.03& 0.02&-0.02& 2.72&-0.03& 0.03&-0.02& 2.72
 \\
 Nb$_1$ &  7.64  &  8.76  &  8.76 &  8.76 & 8.27 & 0.19& 0.19& 8.27&
 6.93& 0.17&-0.13& 0.13& 6.93& 0.17&-0.17& 0.13& 6.93 \\
 Nb$_2$ &  7.64  &  8.75  &  8.75 &  8.76 & 8.27 & 0.19& 0.19&
 8.27& 6.91& 0.19&-0.14& 0.14& 6.91& 0.19&-0.19& 0.14& 6.91
 \\
 O$_1$  & -4.62  & -2.15  & -2.15 & -3.04 &-4.15 &-0.08&-0.29&-2.60&
 -2.33&-0.06&-0.12&-0.08&-2.34& 0.17& 0.23&-0.16&-5.79 \\
 O$_2$  & -5.18  & -2.75  & -2.75 & -3.04 &-2.58
 &-0.29&-0.11&-4.15&-5.79&-0.23& 0.16& 0.12&-2.33&-0.06&-0.17&-0.08&-2.34
 \\
 O$_3$  & -4.62  & -2.15  & -2.15 & -1.97 &-7.05 &-0.50& 0.16&-2.01&
 -2.34& 0.17& 0.08& 0.16&-5.79&-0.23& 0.06& 0.12&-2.33 \\
 O$_4$  & -2.91  & -1.76  & -3.99 & -1.97 &-2.01 &
 0.16&-0.50&-7.05&-3.88& 0.07&-0.13&-0.17&-2.54& 0.04& 0.12& 0.02&-2.56
 \\
 O$_5$  & -2.91  & -3.99  & -1.76 & -3.04 &-2.60 &-0.29&-0.08&-4.15&
 -2.54& 0.04& 0.17& 0.02&-2.56&-0.12&-0.07& 0.13&-3.88 \\
 O$_6$  & -1.94  & -6.84  & -2.18 & -3.04 &-4.15
 &-0.11&-0.29&-2.58&-2.56&-0.12&-0.02& 0.13&-3.88& 0.07&-0.04&-0.17&-2.54
 \\
 O$_7$  & -1.94  & -2.18  & -6.84 & -1.78 &-3.59 & 0.55& 0.55&-3.59&
 -3.86& 0.06&-0.12&-0.15&-2.54& 0.05& 0.11& 0.03&-2.56 \\
 O$_8$  & -1.94  & -6.84  & -2.18 & -6.74 &-2.05
 &-0.05&-0.05&-2.05&-2.54& 0.05& 0.15& 0.03&-2.56&-0.11&-0.06& 0.12&-3.86
 \\
 O$_9$  & -1.94  & -2.18  & -6.84 & -6.74 &-2.05 &-0.05&-0.05&-2.05&
 -2.56&-0.11&-0.03& 0.12&-3.86& 0.06&-0.05&-0.15&-2.54 \\
 Pb$_1$ &  3.43  &  3.80  &  3.80 &  3.96 & 3.97 &-0.27&-0.27&
 3.97&3.85&-0.18& 0.12&-0.12& 3.85&-0.18& 0.18&-0.12& 3.85
 \\
 Pb$_2$ &  3.33  &  3.90  &  3.90 &  3.97 & 3.56 & 0.32& 0.32& 3.56&
 3.79& 0.41&-0.51& 0.51& 3.79& 0.41&-0.41& 0.51& 3.79 \\
 Pb$_3$ &  3.40  &  3.80  &  3.80 &  3.97 & 3.56 & 0.32& 0.32&
 3.56& 3.88&-0.16& 0.10&-0.10& 3.88&-0.16& 0.16&-0.10& 3.88

 \end{tabular} \label{Berry}\end{ruledtabular} \end{table*}

\begin{table*}[!htbp]
\begin{ruledtabular}
 \caption{Diagonal elements of the dynamical matrix (in cm$^{-1}$) for structurally
 stable $[001]_{NNM}$, $[110]_{NNM}$~ and $[111]_{NNM}$~ structures}
\begin{tabular}{c|c|ccccccccccccccc}
 structure&axis&Mg & Nb$_1$ & Nb$_2$ & O$_1$ & O$_2$ & O$_3$ & O$_4$ & O$_5$ &
O$_6$ & O$_7$ & O$_8$ & O$_9$ & Pb$_1$ & Pb$_2$ & Pb$_3$
\\ \hline
$[001]_{NNM}$&$z$&268&332&339&725&585&724&335&343&296&316&255&322&120&106&120\\
 &$x$&379&224&224&254&285&251&492&300&597&264&601&263&62&80&62\\
 &$y$&410&145&145&234&288&233&274&533&230&525&230&526&50&54&49\\
 \hline
$[110]_{NNM}$&$z$&347&233&233&291&293&253&253&291&293&465&604&604&67&72&72\\
&$x$&329&283&286&613&310&654&263&296&602&314&263&262&97&84&87\\
&$y$&329&286&283&296&602&263&654&613&310&314&262&263&97&87&84\\
\hline
$[111]_{NNM}$&$z$&358&280&280&249&602&249&646&277&240&648&239&279&73&67&72\\
 &$x$&331&314&283&270&277&605&264&274&722&275&247&584&84&78&77\\
 &$y$&330&282&314&605&275&268&277&583&251&265&723&277&78&78&84
  \end{tabular} \label{diag} \end{ruledtabular}\end{table*}

\begin{table}[!htbp]\begin{ruledtabular}
 \caption{The ionic coordinates (in $\AA$) for the 30-ion supercells}
\begin{tabular}{c|ccc|ccc}
&\multicolumn{3}{c|}{$[001]_{NNM}$}&\multicolumn{3}{c}{$[111]_{NT}$}
\\ \hline type& $x$  & $y$ & $z$ &$x$ & $y$ & $z$
\\ \hline
  Mg$_1$&    7.98& -0.02&  8.16& -0.07 & 7.95 &-4.06\\
  Mg$_2$&    4.00& -0.00&  4.08&  7.95 &-0.07 &-4.07\\
  Nb$_1$&    4.01& -0.09&  8.11&  3.95 & 3.96 &-4.00\\
  Nb$_2$&    3.95& -0.04&  0.08&  3.96 & 3.96 &-8.02\\
  Nb$_3$&    7.93& -0.02&  0.00&  7.94 & 3.91 &-4.08\\
  Nb$_4$&    3.99& -3.98&  4.12&  3.89 & 7.95 &-4.08\\
  O$_1$&     2.04&  0.40&  8.10&  8.21 & 3.90 &-6.08\\
  O$_2$&     5.96& -0.31&  8.04&  3.82 & 4.15 &-6.10\\
  O$_3$&     4.40&  1.98&  8.12&  3.91 & 0.20 &-2.15\\
  O$_4$&     3.66& -1.98&  8.10&  4.27 & 7.82 &-6.08\\
  O$_5$&     2.05&  0.17&  0.00& -0.12 & 4.18 &-2.14\\
  O$_6$&     6.03& -0.03& 12.02&  4.15 & 3.84 &-2.10\\
  O$_7$&     4.18&  2.01& 12.12&  1.99 & 7.85 &-4.32\\
  O$_8$&     4.00& -1.98& 12.07&  5.95 & 3.82 &-4.32\\
  O$_9$&     1.98&  0.00&  3.86&  5.94 & 0.29 &-4.01\\
  O$_{10}$&  6.06&  0.06&  3.96&  0.14 &10.02 &-4.05\\
  O$_{11}$&  4.02&  2.06&  3.87&  4.22 & 5.96 &-4.03\\
  O$_{12}$&  4.06& -2.02&  3.92&  3.80 & 2.03 &-4.32\\
  O$_{13}$&  0.27& -0.01& 10.21&  2.03 & 4.21 &-3.97\\
  O$_{14}$&  4.00& -3.92&  1.94& -0.11 & 5.94 &-4.37\\
  O$_{15}$&  0.11& -0.04&  6.09& 10.01 &-0.19 &-4.32\\
  O$_{16}$&  3.97&  0.14& 10.11& -2.07 & 8.25 &-4.03\\
  O$_{17}$&  4.12&  0.02&  1.95&  8.21 & 1.99 &-4.00\\
  O$_{18}$&  3.95&  0.20&  6.13&  7.84 &-2.07 &-4.36\\
  Pb$_1$ &   5.75&  1.99& 10.07&  5.91 & 1.70 &-6.02\\
  Pb$_2$&    5.96&  1.98&  2.41&  1.69 & 5.91 &-6.03\\
  Pb$_3 $&   5.84&  1.87&  6.30&  1.95 & 1.92 &-1.76\\
  Pb$_4$&    1.89&  1.74& 10.10&  5.85 & 5.96 &-6.11\\
  Pb$_5$&    5.95& -2.04&  2.40&  5.85 & 1.77 &-1.90\\
  Pb$_6$&    2.06&  1.83&  6.27&  1.84 & 5.89 &-1.83
 \end{tabular}\label{poscar30} \end{ruledtabular}  \end{table}

 \begin{table}[!htbp]
\begin{ruledtabular}
 \caption{The dynamical charges of ions in [001]$_{NCC'}$}
\begin{tabular}{c|ccc|c|ccc}
ion $i$ & $Z^*_{izz}$ & $Z^*_{ixx}$ & $Z^*_{iyy}$ &ion $i$ &
$Z^*_{izz}$ & $Z^*_{ixx}$ & $Z^*_{iyy}$ \\ \hline
 Mg$_1$  & 2.74 & 2.65 & 2.65 &
 Mg$_2$  & 2.74 & 2.65 & 2.65 \\
 Nb$_1$  & 6.48 & 6.00 & 6.00 &
 Nb$_2$  & 7.87 & 9.11 & 9.11 \\
 Nb$_3$  & 7.87 & 9.11 & 9.11 &
 Nb$_4$  & 6.48 & 6.00 & 6.00 \\
 O$_1$   &-2.61 &-3.62 &-2.82 &
 O$_2$   &-2.61 &-3.62 &-2.82 \\
 O$_3$   &-2.61 &-2.82 &-3.62 &
 O$_4$   &-2.61 &-2.82 &-3.62 \\
 O$_5$   &-1.99 &-7.04 &-2.09 &
 O$_6$   &-1.99 &-7.04 &-2.09 \\
 O$_7$   &-1.99 &-2.09 &-7.04 &
 O$_8$   &-1.99 &-2.09 &-7.04 \\
 O$_9$   &-2.61 &-3.62 &-2.82 &
 O$_{10}$&-2.61 &-3.62 &-2.82 \\
 O$_{11}$&-2.61 &-2.82 &-3.62 &
 O$_{12}$&-2.61 &-3.62 &-2.82 \\
 O$_{13}$&-4.15 &-2.57 &-2.57 &
 O$_{14}$&-5.71 &-2.42 &-2.42 \\
 O$_{15}$&-3.93 &-2.44 &-2.44 &
 O$_{16}$&-5.71 &-2.42 &-2.42 \\
 O$_{17}$&-4.15 &-2.57 &-2.57 &
 O$_{18}$&-3.93 &-2.44 &-2.44 \\
 Pb$_1$  & 3.52 & 4.41 & 4.41 &
 Pb$_2$  & 3.52 & 4.41 & 4.41 \\
 Pb$_3$  & 4.17 & 3.93 & 3.93 &
 Pb$_4$  & 3.52 & 4.41 & 4.41 \\
 Pb$_5$  & 3.52 & 4.41 & 4.41 &
 Pb$_6$  & 4.17 & 3.93 & 3.93
 \end{tabular} \label{Berry30}\end{ruledtabular} \end{table}

\begin{table}[!htbp]
\begin{ruledtabular}
 \caption{The diagonal frequencies for the 30-ion supercells of PMN}
\begin{tabular}{c|ccc|ccc}
&\multicolumn{3}{c|}{[001]$_{NCC'}$} &
\multicolumn{3}{c|}{[111]$_{NT}$}
\\ \hline
 ion & $z$ & $x$ & $y$ & $z$ & $x$ & $y$
\\ \hline
 Mg$_1$  &346 & 328 & 340 &    311 & 334 & 354 \\
 Mg$_2$  &276 & 340 & 344 &    308 & 335 & 284\\
 Nb$_1$  &279 & 271 & 290 &    289 & 288 & 279\\
 Nb$_2$  &322 & 276 & 270 &    272 & 298 & 307\\
 Nb$_3$  &318 & 286 & 251 &    302 & 276 & 269\\
 Nb$_4$  &301 & 288 & 283 &    295 & 296 & 291\\
 O$_1$   &385 & 537 & 333 &    585 & 271 & 277\\
 O$_2$   &262 & 608 & 251 &    675 & 266 & 266\\
 O$_3$   &399 & 331 & 521 &    618 & 274 & 256\\
 O$_4$   &256 & 267 & 648 &    548 & 284 & 299\\
 O$_5$   &300 & 640 & 248 &    626 & 268 & 247\\
 O$_6$   &263 & 657 & 257 &    682 & 281 & 276\\
 O$_7$   &297 & 272 & 625 &    284 & 670 & 255\\
 O$_8$   &281 & 246 & 633 &    273 & 636 & 261\\
 O$_9$   &292 & 577 & 302 &    314 & 587 & 296\\
 O$_{10}$ & 281 & 612 & 278&   266 & 273 & 646\\
 O$_{11}$ & 285 & 278 & 600 &  273 & 268 & 659\\
 O$_{12}$ & 279 & 299 & 613 &  287 & 275 & 625\\
 O$_{13}$ & 555 & 264 & 294 &  281 & 662 & 265\\
 O$_{14}$ & 661 & 267 & 261 &  292 & 295 & 639\\
 O$_{15}$ & 735 & 214 & 207 &  290 & 602 & 324\\
 O$_{16}$ & 557 & 241 & 250 &  284 & 628 & 249\\
 O$_{17}$ & 733 & 233 & 253 &  302 & 295 & 630\\
 O$_{18}$ & 596 & 244 & 245 &  276 & 310 & 631\\
 Pb$_1$  & 108 & 88  & 76  &    81 &  81 &  92\\
 Pb$_2$  & 81  & 78  & 81  &    87 &  86 &  70\\
 Pb$_3$  & 93  & 81  & 68  &    97 &  76 &  94\\
 Pb$_4$  & 105 & 76  & 89  &    78 &  78 &  81\\
 Pb$_5$  & 88  & 79  & 70  &    84 &  82 &  85\\
 Pb$_6$  & 92  & 67  & 76  &    86 &  83 &  76\\
 \end{tabular} \label{diag30}\end{ruledtabular} \end{table}

 \begin{table*}[!htbp]
\begin{ruledtabular}
 \caption{Assignment of the frequencies obtained for tetragonally averaged
[PMN]$_{NCC'}$: $n$~ is the degeneracy, $\alpha$~ direction of
displacements, {\bf{q}} the wave vector corresponding to the
reduced primitive unit cell, $\varepsilon$~ the averaged over the
directions contribution to dielectric permittivity. Notice that
these frequencies differ from those obtained in the straight
computations and, hence, can be used only for a qualitative
analysis.}
\begin{tabular}{cccccc}
 n & $\alpha$ & $\nu$ & \bf{q}  & type & content ($\varepsilon$)\\ \hline
 3 & xyz &   0 & $(0,0,0)$               & acoustic   & all ions \\
 4 & xy  &  61 & $(0,0,2\pi/3a)$         & acoustic   & Pb       \\
 1 &  z  &  68 & $(\pi/a,\pi/a,0)$       & acoustic   & Pb       \\
 2 &  z  &  83 & $(\pi/a,\pi/a,2\pi/3a)$ & acoustic   & Pb       \\
 4 & xy  &  85 & $(\pi/a,\pi/a,2\pi/3a)$ & acoustic   & Pb       \\
 2 & xy  &  94 & $(\pi/a,\pi/a,0)$       & acoustic   & Pb       \\
 2 & xy  &  97 & $( 0, 0, 0 )$           & optical, Last-type &
                                           Pb-BO$_6$~ translation (11.3) \\
 2 & z   & 111 & $(0,0,2\pi/3a)$         & acoustic   & Pb       \\
 1 & z   & 121 & $(0,0,0)$               & optical, Last-type &
                                           Pb-BO$_6$~ translation (5.3)  \\
 1 & z   & 178 & $(\pi/a,\pi/a,0)$       & optical    &B, O$_z$  \\
 4 & xy  & 192 & $(0,0,2\pi/3a)$         & optical    &B, O      \\
 2 & xy  & 232 & $(\pi/a,\pi/a,0)$       & optical    &B, O$_z$    \\
 4 & xy  & 249 & $(0,0,2\pi/3a)$         & optical    &B, O      \\
 2 & xy  & 249 & $(0,0,0)$               & optical, polar    &B-O$_z$~bending
                                                            (0.9)\\
 4 & xy  & 252 & $(\pi/a,\pi/a,2\pi/3a)$ & optical    &B, O      \\
 2 & xy  & 253 & $(0,0,0)$               & optical, polar    &B-O$_{xy}$~ bending
                                                            (4.2)\\
 1 & z   & 253 & $(0,0,0)$               & optical, nonpolar & O$_{xy}$~ bending \\
 4 & xy  & 265 & $(0,0,2\pi/3a)$         & optical    &B, O      \\
 1 & z   & 266 & $(0,0,0)$               & optical    &B-O$_{xy}$~ bending (4.3) \\
 2 & z   & 266 & $(0,0,2\pi/3a)$         & optical    &O$_{xy}$      \\
 4 & z   & 270 & $(\pi/a,\pi/a,2\pi/3a)$ & optical    &O$_{xy}$      \\
 2 & z   & 284 & $(\pi/a,\pi/a,2\pi/3a)$ & optical    &B, O$_z$    \\
 2 & z   & 287 & $(0,0,2\pi/3a)$         & optical    &B, O      \\
 2 & z   & 292 & $(\pi/a,\pi/a,0)$       & optical    &O$_{xy}$      \\
 4 & xy  & 292 & $(\pi/a,\pi/a,2\pi/3a)$ & optical    &O$_{xy}$      \\
 2 & xy  & 318 & $(\pi/a,\pi/a,0)$       & optical    &O$_{xy}$      \\
 4 & xy  & 342 & $(\pi/a,\pi/a,2\pi/3a)$ & optical    &B, O$_z$    \\
 2 & xy  & 354 & $(\pi/a,\pi/a,0)$       & optical    &B, O$_z$    \\
 2 & z   & 402 & $(0,0,2\pi/3a)$         & optical    &B, O      \\
 1 & xy  & 571 & $(0,0,2\pi/3a)$         & optical    &B, O      \\
 2 & xy  & 591 & $(0,0,0)$               & optical, polar    &B-O stretching (0.8) \\
 1 &  z  & 599 & $(\pi/a,\pi/a,0)$       & optical    &B, O$_z$    \\
 1 &  z  & 606 & $(0,0,0)$               & optical, polar    &B-O stretching (0.7) \\
 2 &  z  & 660 & $(\pi/a,\pi/a,2\pi/3a)$ & optical    &B, O$_z$    \\
 4 & xy  & 677 & $(\pi/a,\pi/a,2\pi/3a)$ & optical    &O$_{xy}$      \\
 2 & xy  & 694 & $(\pi/a,\pi/a,0)$       & optical    &O$_{xy}$      \\
 2 & z   & 729 & $(0,0,2\pi/3a)$         & optical    &B, O
 \end{tabular} \label{assignment} \end{ruledtabular}\end{table*}

\begin{table}[!htbp]
\begin{ruledtabular}
 \caption{Eigen vectors for the zone-center modes polarized in the
 z direction obtained with the help of the tetragonally averaged
 dynamical matrix of [001]$_{NCC'}$}
\begin{tabular}{c|ccccc}
 $\nu$ & B & O$_x$ & O$_y$ & O$_z$ & Pb \\ \hline
 121 & 0.027 & 0.035 & 0.035 & 0.032 & -0.017 \\
 253 & 0    & -0.072 & 0.072 & 0 & 0 \\
 266 & 0.031& -0.056 & -0.056 & -0.003 & -0.002 \\
 606 & -0.014 & -0.019 & -0.019 & 0.094 & 0.000
 \end{tabular} \label{eigen} \end{ruledtabular}\end{table}

\begin{table}[!htbp]
 \caption{The Wyckoff positions in the $Immm$~ group.}
\begin{tabular}{c|c}
a position & ions
\\ \hline
 4f & Mg$_1$,  Mg$_2$\\
 2c & Nb$_1$\\
 2a & Nb$_2$\\
 4e & Nb$_3$, Nb$_4$\\
 8m & O$_1$, O$_3$, O$_4$, O$_5$ \\
 4i & O$_2$, O$_6$\\
 4n & O$_7$, O$_9$, O$_{14}$, O$_{17}$ \\
 4n & O$_8$, O$_{11}$, O$_{12}$, O$_{13}$\\
 4n & O$_{10}$, O$_{15}$, O$_{16}$, O$_{18}$ \\
 8m & Pb$_1$, Pb$_2$, Pb$_5$, Pb$_6$\\
 4j & Pb$_3$, Pb$_4$
 \end{tabular} \label{Wyckoff}
 \end{table}

Ordered supercells with PMN stoichiometry require 15n atoms (n =
1, 2...). The calculations presented here are for three 15 atom
cells, and two 30 atom cells (Fig.~\ref{structures}): a)
[001]$_{NNM}$; b) [110]$_{NNM}$; c) [111]$_{NNM}$; d)
[001]$_{NCC'}$; e) [111]$_{NT}$. Here: subscript N = a Nb-layer
parallel to  $(h,k,l)$; M = a Mg-layer parallel to $(h,k,l)$; C =
a chessboard ordered layer, i.e. a Mg$_{1/2}$Nb$_{1/2}$-layer or
(001)$_{1:1}$-layer; C$^{\prime}$ = a  chessboard ordered
(001)$_{1:1}$ -layer that is displaced by [1/2,0,0] relative to an
adjacent C-layer; T = a (111)$_{2:1}$-layer of composition
Mg$_{2/3}$Nb$_{1/3}$, which is triangularly-ordered (Fig.
\ref{triangles}a). The [001]$_{NCC'}$~ structure (d) has
(001)$_{Nb}$~ layers that alternate with $CC'$~ double layers
(NaCl-type blocks). This structure was proposed \cite{Burton97} as
a possible cation ordering GS for PMN, however the stability
analyses reported below indicate that the [111]$_{NT}$~ structure
is lower in energy. The [111]$_{NT}$ structure has (111)$_{Nb}$
layers alternating with (111)$_{Mg_{2/3}Nb_{1/3}}$ layers, and, as
such, is an ordered approximate of the 1:1 ``random-site model".
Although labeled ``NT" for Nb layer/triangular layer alternation,
the cation arrangement in the 2:1 layers depends on the layer's
orientation. In Fig. \ref{structures}e, there are
triangularly-ordered layers perpendicular to [111] and
[$\overline{1} \overline{1} 1$] and striped layers (Fig. 3b)
perpendicular to [$1 \overline{1} 1$] and [$\overline{1}11$].

For each ordered supercell, two space group symmetries, and two
corresponding energies, are given in Table~\ref{symmetry}:
$G_{Chem}$~ indicates the space group symmetry dictated by
chemical ordering without additional displacive instabilities;
$G_{SSS}$~ is space group of the lower symmetry ``structurally
stable state." The SSS was determined by applying random
perturbations to the atomic positions, and then relaxing the
system until the absolute values of all force constants are less
than 0.001 eV/\AA$^2$. The difference $\Delta E_{SSS} - \Delta
E_{Chem}$~ is referred to as the relaxation energy.

In general, $G_{SSS} \subset G_{Chem} \subset Pm\overline{3}m$,
and $\Delta E_{SSS} < \Delta E_{Chem}$.  Randomness of
the initial perturbation does not guarantee that the SSS
is in fact the ground-state of the system($G_{SSS}=G_{GS}$), but
it does guarantee that it is, at least, metastable.

We show here more tables (see Tables \ref{poscar} to
\ref{Wyckoff}) than in our submitted paper, in order to save place
in the journal.

\subsection{Short-Range Order and Lattice Averaged Spin Products}

Chemically ordered, disordered and random configurations can be quantitatively analyzed
with respect to their short-range order (SRO) correlations,
and one can answer such questions as:
which configuration is most similar to the random state \cite{Wei,Zunger91a,Zunger91b}
or, more relevant for this work, which is most similar to the
random site model (RSM)?
The procedure is to calculate a lattice averaged spin product, $\overline{\prod}(r,t)_{S}$,
for each structure (S) and cluster $(r,t)$:

\begin{eqnarray}
\overline{\prod}(r,t)_{S} = M(0) \xi(0)_{S}  +  M(1) \xi(1)_{S} +
\nonumber \\ M(2,1) \xi(2,1)_{S} ...
       ... + M(8) \xi(8)_{S}
\end{eqnarray}
Here, $\overline{\prod}(r,t)_{S}$~ is the product between
multiplicities, $M(r,t)$, for r-body clusters of type $t$~
and correlation functions, $\xi(r,t)$, for the clusters.
The parameter r=1,2,..., is the number of sites in the cluster
(1=site, 2=pair ...8=cube), and $t$~ is an arbitrary index (e.g. t=2 for
an r-body cluster of type 2).
In a two component system, there is a one to one correspondence between
symmetrically distinct clusters
and $(r,t)$~ clusters (symmetrically distinct in the high-symmetry phase).
The correlation function for any cluster is
$\xi(r,t) = < \sigma_i \sigma_j ... \sigma_r >$,
where: $<\sigma_i \sigma_j ... \sigma_r >$~ is an ensemble average;
$\sigma = 1$~ for Nb; and $\sigma = -1$~ for Mg.
In general: $\xi(0)=1$; $-1 \leq ~\xi(1)~ \leq 1$~ for the site (point) correlation;
$-1~\leq~\xi(2,1)~ \leq 1$~ for the first nn pair; $-1 \leq ~\xi(2,2)~
\leq 1$~ for the 2nd nn pair, etc.
The $\{\xi(r,t)\}$~ sets listed in Table \ref{tb:Ztable} are
truncated at the cube-approximation; i.e. only the correlations
for cubes and their subclusters, are included.
Indices in column 1 identify symmetrically distinct
subclusters of the cube in space group $P\overline{3}m$,
labeled as in figure~\ref{oxygens}; column 2
lists the $M(r,t)$; columns 3-9 list the $\xi(r,t)$~ for the
random state, RSM, and the $[001]$, $[111]$, $[110]$, and
$[111]_{NCC'}$ supercells, respectively. \\

\begin{figure}[!htbp]
\resizebox{0.4\textwidth}{!} {\includegraphics{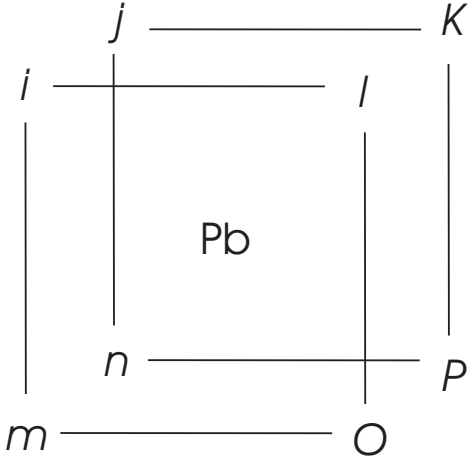}}
\caption{8-cite configuration. } \label{oxygens}
\end{figure}

The $\{\xi(r,t)\}$~ sets for the $[001]$, $[111]$,
$[110]$, and $[111]_{NCC'}$ structures were calculated by counting
the distributions of distinct cube-cluster configurations
in each structure, and the distribution of
subclusters therein.  Correlation functions for the random
configuration were calculated using

\begin{eqnarray}
&&x(8)_{ijklmnop} = \nonumber \\
&&=x(1)_{i}x(1)_{j}x(1)_{k}x(1)_{l}x(1)_{m}x(1)_{n}x(1)_{o}x(1)_{p},
\end{eqnarray}
where $x(8)_{ijklmnop}$~ is the probability that a cube (8-body
cluster) has configuration $ijklmnop$, and $x(1)_{i}$~ is the
probability of finding atom-$i$ on the site labeled $i$~ in Figure
~\ref{oxygens}. That is, $x(1)_{i}=2/3$~ when Nb occupies
site-$i$, and $x(1)_{i}=1/3$~ when Mg occupies site-$i$. Cluster
algebra [relationships between the $x(r,t)$~ and $\xi(r,t)$] for a
two component system was fully described by Sanchez and de
Fontaine \cite{Sanchez}, and the extension to multicomponent
systems was described in Sanchez et. al.  \cite{Sanchez84}.

The set $\{\xi(r,t)\}$, for the RSM, was approximated numerically by
generating a 399x399x399 site
simulation box with ideal NaCl-type ordering of Nb and Mg, randomly changing 1/3 of
the Mg ions to Nb, and counting cube-, and cube-subcluster, correlations
in the simulation box.

Because $\overline{\prod}(r,t)_{S}$~ for any ordered state or random ensemble is a vector, one
can unambiguously (except for truncation errors) compare average
Pythagorean distances between $\overline{\prod}(r,t)_{S}$~ for different
states of ordered. The last two rows of Table \ref{tb:Ztable} list average
Pythagorean distances relative to the random configuration,$R_{S-Random}^{\bullet}$,
and the RSM, $R_{S-RSM}$:

\begin{eqnarray}
R_{S-RSM} = \frac{1}{22} \{~[~M(  0)( \xi(0)_{S}   - \xi(0)_{RSM}  )~]^{2}    \nonumber \\
             +~[~M(  1)( \xi(1)_{S}   - \xi(1)_{RSM}  )~]^{2}    \nonumber \\
             +~[~M(2,1)( \xi(2,1)_{S} - \xi(2,1)_{RSM})~]^{2}    \nonumber \\
             +~[~M(2,2)( \xi(2,2)_{S} - \xi(2,2)_{RSM})~]^{2}              \\
             +~[~M(3,1)( \xi(3,1)_{S} - \xi(3,1)_{RSM})~]^{2}    \nonumber \\
             +~...~~~~~~~~~~~~~~~~~~~~~~~~~~~~~~~~~~~~~~~~~~     \nonumber \\
             +~[~M(8)( \xi(8)_{S} - \xi(8)_{RSM})~]^{2}~~\}^{1/2} \nonumber
\end{eqnarray}

A smaller value of $R_{S-RSM}$~
implies a structure, S, that is more similar to the RSM, so
clearly the $[111]_{NT}$~ is a better approximate of the RSM
than any of the other supercells listed in Fig. \ref{structures}.

\begin{table*} [!ht]
\caption{Multiplicities and correlation functions for: a random
distribution; the random-site model (RSM); the five ordered
supercells. }
\begin{ruledtabular}
\begin{tabular}{|c|c|c|c|c|c|c|c|c|}
  {Cluster }    & {$M(r,t)^{\dag}$} & {$Random$} & {$RSM^{\ddag}$} & {$[001]$} & {$[111]$}
    & {$[110]$} & {$[111]_{NCC'}$} & {$[111]_{NT}$} \\ \hline
  {0$^{\star}$} &  {1} &  {1}       & { 1.00000} &  {1}   & {1}    & {1}    & {1}    & {1}    \\ \hline

      {i}       &  {1} & {-1/3}     & {-0.33330} & {-1/3} & {-1/3} & {-1/3} & {-1/3} & {-1/3} \\ \hline

      {ij}      &  {3} & { 1/9}     & { 0.10686} & { 5/9} & {-1/3} & { 1/9} & {-1/3} & {-1/9} \\
      {ik}      &  {6} & { 1/9}     & { 0.02985} & { 1/9} & { 1/3} & {-1/9} & { 5/9} & { 0  } \\
      {io}      &  {4} & { 1/9}     & { 0.06944} & {-1/3} & { 0  } & { 1/3} & {-1/3} & { 1/3} \\  \hline

      {ijk}     & {12} & {-1/27}    & { 0.03190} & {-1/3} & { 1/3} & {-1/9} & { 1/9} & { 2/9} \\
      {ikn}     &  {8} & {-1/27}    & { 0.05100} & {-1/3} & {-1/3} & { 1/3} & {-1/3} & { 0  } \\
      {ijo}     & {24} & {-1/27}    & {-0.00544} & {-1/3} & { 0  } & { 1/9} & { 1/9} & { 0  } \\ \hline

     {ijkl}     &  {3} & { 1/81}    & { 0.05259} & { 1}   & {-1/3} & { 5/9} & { 1/9} & {-1/9} \\
     {ikmo}     & {12} & { 1/81}    & { 0.05528} & { 1}   & { 1/3} & { 1/9} & { 1/9} & {-1/9} \\
     {ijlm}     &  {8} & { 1/81}    & {-0.04396} & {-1/3} & {-1/3} & { 1/3} & {-1/3} & { 0  } \\
     {iknp}     &  {1} & { 1/81}    & { 0.00030} & { 1}   & {-1/3} & {-1/3} & { 1  } & { 1/3} \\
     {ijlo}     & {24} & { 1/81}    & {-0.03173} & { 5/9} & { 0  } & {-1/3} & {-1/3} & { 0  } \\
     {ijko}     & {48} & { 1/81}    & {-0.07826} & { 1/9} & { 0  } & {-1/9} & { 1/9} & {-1/9} \\ \hline

    {ijklm}     & {24} & {-1/243}   & {-0.02746} & {-1/3} & { 0  } & {-1/9} & { 1/9} & {-1/9} \\
    {ijknp}     &  {8} & {-1/243}   & {-0.00626} & {-1/3} & { 2/3} & { 1/3} & {-1/3} & {-1/3} \\
    {ijkmo}     & {24} & {-1/243}   & { 0.09270} & {-1/3} & {-1/3} & { 1/9} & { 1/9} & { 1/9} \\ \hline

   {ijklmn}     & {12} & { 1/729}   & { 0.06074} & { 5/9} & { 1/3} & { 1/9} & {-1/3} & { 1/9} \\
   {ijklmo}     & {12} & { 1/729}   & {-0.00138} & { 1/9} & {-1/3} & {-1/9} & { 5/9} & { 2/9} \\
   {jklmnp}     &  {8} & { 1/729}   & {-0.28448} & {-1/3} & { 2/3} & { 1/3} & {-1/3} & {-1/3} \\ \hline

  {ijklmno}     &  {8} & {-1/2187}  & { 0.01467} & {-1/3} & { 0  } & {-1/3} & {-1/3} & { 0  } \\ \hline

 {ijklmnop}     & {1}  & { 1/6561}  & {-0.04489} & { 1}   & {-1/3} & { 1  } & { 1  } & {-1/3} \\ \hline \hline
{$R_{S-Random}^{\bullet}$} &  &   {0}   &  {0.776} & {1.165} & {0.653} & {0.582} & {0.690} & {0.401} \\ \hline
{$R_{S-RSM}$             } &  & {0.776} &   { 0}   & {1.297} & {1.006} & {1.080} & {1.201} & {0.747} \\
\end{tabular}
\end{ruledtabular}
\label{tb:Ztable} \begin{flushleft} $^{\dag}$~ M(r,t) is the
multiplicity, per site, of the r-body cluster (r=1,2,...8 sites)
of type t. \\ $^{\ddag}$~ Real numbers are from a 399x399x399 site
simulation, and fractions are inferred, from multiple simulations.
\\ $^{\star}$~ The Zero- or empty cluster. \\ $^{\bullet}$~
$R_{S-Random}$~ is the average (divided by the number of
correlation functions, 22) Pythagorean distance between
the LASP of structure S, and the LASP for a random alloy; and
similarly for $R_{S-RSM}$. \end{flushleft}
\end{table*}

\section{Structurally Stable States and Lattice Dynamics}

Lattice dynamical force constants were calculated by the frozen
phonon method; i.e. in turn, displacing each ion by 0.01~\AA~ in
each of the three orthogonal (Cartesian) directions; the force
matrix is $R_{i\alpha,j\beta}$; where $i$~ and $j$~ index ionic
positions, $\alpha$~ and $\beta$~ are the Cartesian directions.
The dynamical matrix DM is
$D_{i\alpha,j\beta}=R_{i\alpha,j\beta}/\left(M_iM_j\right)^{1/2}$,
where $M_i$~ is the $i$-th ionic mass. Vibrational mode
frequencies were found by diagonalizing the DM, and used to
compute IR reflectivity spectra (2):

\begin{eqnarray}
&& \varepsilon_{\alpha\beta}(\nu) = \varepsilon_{\infty \alpha
\beta} + \sum_i\frac{\varepsilon_{i\alpha \beta}
\nu_i^2}{\nu_i^2-\nu^2+i\gamma_i\nu}
\\
&&\varepsilon_{i \alpha \beta} =
\frac{Z^*_{i\alpha}Z^*_{i\beta}}{4\pi^2V\varepsilon_0m_0\nu^2_i}
\\ &&Z^*_{\mu\alpha} =
\sum_{i\gamma}{Z^*_{i\alpha\gamma}\left(m_0/M_i\right)^{1/2}a_{\mu i\gamma} }
\end{eqnarray}
Here $V$~ is unit cell volume, $M_i$~ is atomic mass, $m_0=1
a.e.m.$, $\varepsilon_\infty=5.83$, the experimental electronic
dielectric permittivity, $\nu$ is the frequency of the incident
radiation, $\nu_i$~ is the frequency of mode $i$, and $a_{\mu i
\gamma}$~ the component of the DM eigen-vector for the $\mu$-th
mode, $i$-th ion, and $\gamma$~ direction. The damping constant
was set to 60 cm$^{-1}$, the approximate average of damping
constants that were fit to experimental results (Table I).

\begin{figure}
\resizebox{0.45\textwidth}{!}{\includegraphics{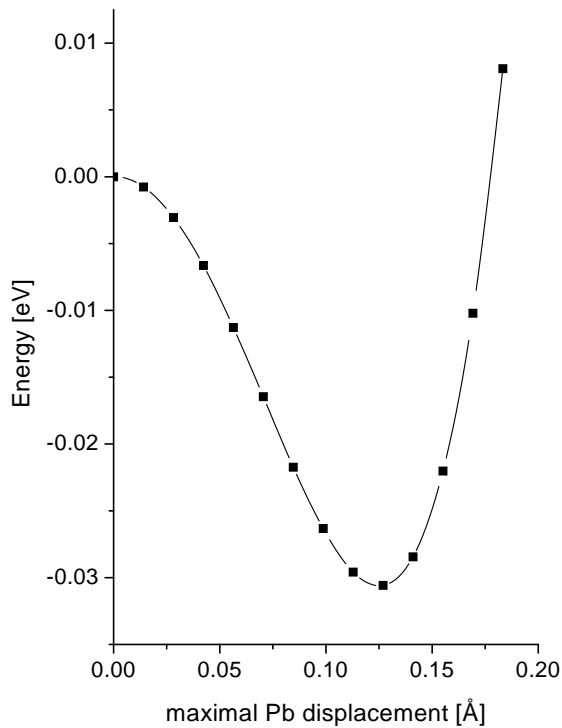}}
\caption{Computed potential relief for the Last-type soft mode in
the $[001]_{NCC'}$~ structure.} \label{barrier}
\end{figure}

\begin{figure}[!htbp]
\resizebox{0.4\textwidth}{!} {\includegraphics{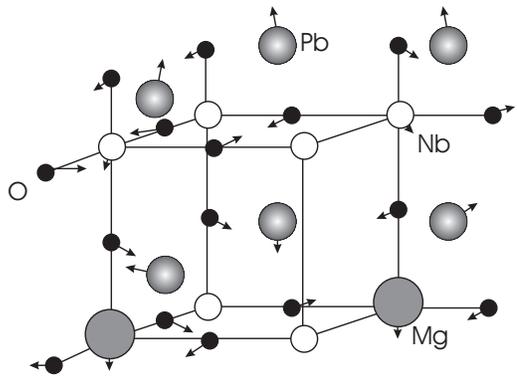}}
\caption{Atomic shifts in the equilibrium structure of
[111]$_{NT}$. The average shift of the Pb ions is approximately
along a $[\overline{bb}c]$ direction.}\label{oxygens}
\end{figure}

\begin{figure}
\resizebox{0.45\textwidth}{!}{\includegraphics{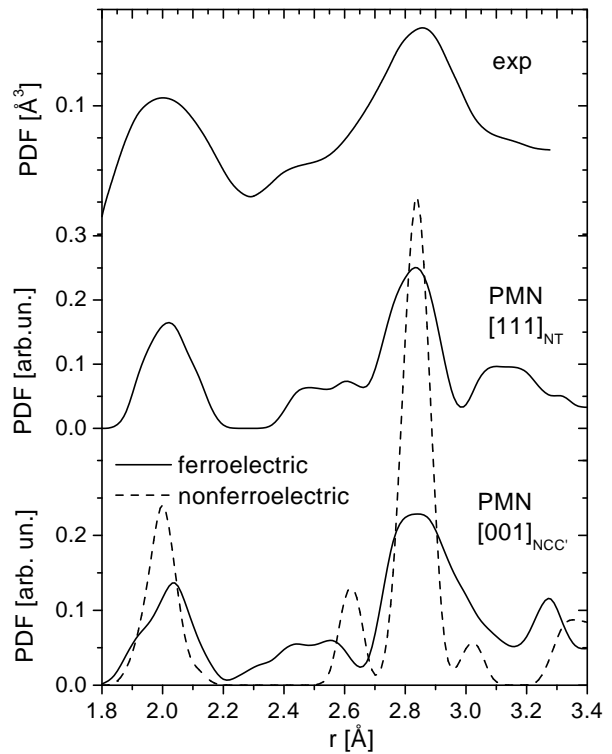}}
\caption{Computed pair distribution function in the
[001]$_{NCC'}$~ and [111]$_{NT}$~structures of PMN. PDF's are
compared with experimental results of Egami.\cite{Egami} }
\label{PDF}
\end{figure}

\subsection{Ionic relaxation}

 We first consider the relaxation of ions with respect to
ideal perovskite positions in the $G_{Chem}$ structures. In every
structure except $[111]_{NNM}$, the largest displacements involve
Pb.  In $[001]_{NNM}$, the Pb-ions with 4 Mg neighbors displace by
0.47 \AA~toward the (001)$_{Mg}$~ layers in which the O-ions are
underbonded (coordinated by two Mg$^2+$~ ions
\cite{Burton99,Burton00MSMSE}). In $[110]_{NNM}$, the Pb-ions with
2 Mg neighbors displace by 0.47 \AA~toward the underbonded O-ion
between the 2 Mg. The other structures do not contain nearest
neighbor Mg-Mg pairs; nonetheless, each Pb-ions in an asymmetric
environment of Mg and Nb is displaced off-center by a nonzero
``local field".\cite{Bur04} The magnitude of these displacements
ranges from 0.25 \AA~ to 0.27 \AA~ for the [001]$_{NCC'}$~ and
[111]$_{NT}$ structures, but is only 0.04 \AA~ for the
[111]$_{NNM}$ structure.  The small Pb displacements in the
[111]$_{NNM}$ structure result because of a special local
environment of Pb in this structure. The largest ionic motion is
by the Nb-ions, which shift 0.09 \AA~towards nearest neighbor
(111)$_{\rm Mg}$ planes. In the [111]$_{NT}$ structure, Pb
deviates from the [111] axis; a net displacement of
(0.17,0.17,0.08) \AA~ occurs.

Next, we consider the nature and result of symmetry-breaking
relaxations leading to the $G_{SSS}$ structures. All the 15-ion
supercell have multiple instabilities. While the different shapes
of the supercells studied prevents direct comparison of the modes
between the different cells, there are both ferroelectric
instabilities of the Last type (Pb motion against the other ions),
and antiferrodistortive (tilting) instabilities dominated by O
motion. Fig. \ref{barrier} shows the potential barrier for the
Last mode in [001]$_{NCC'}$~ structure. With respect to the
primitive perovskite cell, every supercell  shows wide dispersion
of Pb- and O- dominated instabilities across the Brillouin zone.
Nb motion opposite O in Nb-O-Nb... chains and planes also plays a
role in the FE instabilities.  We note, however, that long
Nb-O-Nb... chains and planes are unlikely to form experimentally.

Given the existence of multiple instabilities in PMN supercells,
it is not surprising that the $G_{SSS}$ states involve the
freezing of multiple modes. Figure~\ref{oxygens} shows the
complexity of the local structure in fully relaxed [111]$_{NT}$.
Pair distribution functions (PDF) for the $[001]_{NCC'}$~ in
$P4/nmm$~ (dotted) and $P1$~ (solid) structures are shown in Fig.
\ref{PDF}. The peak at 2.8 \AA~is already split by the different
Pb-displacements in the $G_{Chem}$ structures, and the splitting
is significantly enhanced in the $G_{SSS}$ structures.  The
agreement between the predicted and experimental PDF is quite
good.

We estimate the polarizations of our stable structures by using
the displacements of each ion from its position in $G_{Chem}$, and
estimated Born effective charges determined from averaging the
calculated Born effective charges for the $[001]_{NCC'}$~
structure\cite{Wash00,Will}: Z$^*_{\rm Pb} \approx 4.0$; Z$^*_{\rm
Mg} \approx 2.6$; Z$^*_{\rm Nb} \approx 7.4$; Z$^*_{\rm
O_\parallel} \approx -4.8$; Z$^*_{\rm O_\perp} \approx -2.5$. The
results are shown in Table~\ref{gspol.tbl}. The estimated
polarization magnitudes are similar in magnitude, though about 10
\% smaller than the zero-temperature polarizations estimated from
first principles for PbTiO$_3$\cite{Wag97} and
PbSc$_{1/2}$Nb$_{1/2}$O$_3$.\cite{Coc01} Remarkably, three of the
five structures have SSS polarizations that are pseudomonoclinic
(close to an [aab]-type direction).

\begin{table}[h]
\begin{ruledtabular}
\caption{Estimated polarization (in C/m$^2$) of
structurally stable states of PMN supercells.}
\begin{tabular}{c|cccc}
structure & P$_x$ & P$_y$ & P$_z$ & $| {\rm P} |$ \\
\hline
$[001]_{NNM}$  &  0.525 &  0.191 &  0.000 & 0.558 \\
$[110]_{NNM}$  &--0.158 &--0.158 &--0.547 & 0.591 \\
$[111]_{NNM}$  &  0.474 &  0.476 &  0.016 & 0.672 \\
$[001]_{NCC'}$ &--0.228 &--0.243 &  0.385 & 0.509 \\
$[111]_{NT}$   &--0.238 &--0.221 &  0.346 & 0.474
\end{tabular}
\end{ruledtabular}
\label{gspol.tbl}
\end{table}

\begin{table}[!htbp]\begin{ruledtabular}
 \caption{The ionic coordinates for the structurally stable [111]$_{NT}$~
 supercells.
 In units of the basic vectors.}
\begin{tabular}{c|ccc}
type & 1  & 2 & 3\\ \hline
 Pb & 0.920754 &  0.569936 &  0.214546\\
  & 0.571441 &  0.922035 &  0.212031\\
 & 0.219389 &  0.216641 &  0.269811\\
 & 0.752815 &  0.762316 &  0.731786\\
 & 0.406203 &  0.066264 &  0.724385\\
 & 0.058984 &  0.396617 &  0.746027\\
 Mg &0.169985&  0.836951& 0.489575\\
 & 0.838482&  0.171084&  0.488449\\
 Nb &0.495887&  0.496219& 0.501696\\
 & 0.993701&  0.993477&  0.006489\\
 & 0.674003&  0.338406&  0.986974\\
 & 0.337066&  0.674674&  0.986512\\
 O &0.932277&  0.574068&  0.773252\\
 & 0.741991&  0.769317&  0.249203\\
 & 0.421168&  0.112100&  0.251287\\
 & 0.605392&  0.901242&  0.771526\\
 & 0.086447&  0.444319&  0.246015\\
 & 0.273884&  0.247484&  0.746089\\
 & 0.291333&  0.779392&  0.705622\\
 & 0.623756&  0.446513&  0.697410\\
 & 0.731616&  0.261126&  0.289206\\
 & 0.091774&  0.913712&  0.777444\\
 & 0.427503&  0.571872&  0.783701\\
 & 0.608550&  0.461512&  0.201063\\
 & 0.401129&  0.582570&  0.294687\\
 & 0.290460&  0.793188&  0.193402\\
 & 0.960049&  0.111226&  0.703000\\
 & 0.069719&  0.928613&  0.280924\\
 & 0.755079&  0.237712&  0.789281\\
 & 0.952809&  0.127627&  0.187698
  \end{tabular}\label{poscar30} \end{ruledtabular}  \end{table}

\subsection{IR spectra and lattice dynamics of
relaxed structures}

\begin{figure}
\resizebox{0.45\textwidth}{!}{\includegraphics{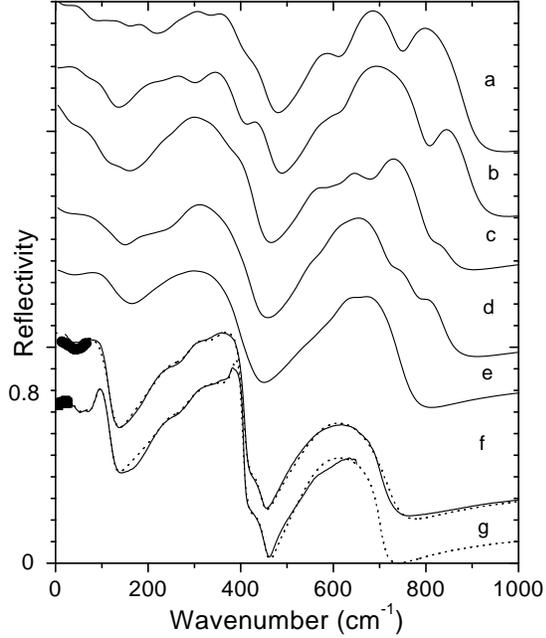}}
\caption{Computed (a-e) and experimental (f-g) IR reflectivity of
PMN: a) [001]$_{NNM}$; b) [011]$_{NNM}$; c) [111]$_{NNM}$; d)
[001]$_{NCC'}$; (e) [111]$_{NT}$; f) 20 K; g) 300 K. Dotted lines
are damped oscillator model fits (see Table \ref{IRmodes}) to the
experimental IR reflectivity spectra (solid lines) and THz spectra
(solid points).}\label{reflectivity}
\end{figure}

\begin{figure}
\resizebox{0.45\textwidth}{!}{\includegraphics{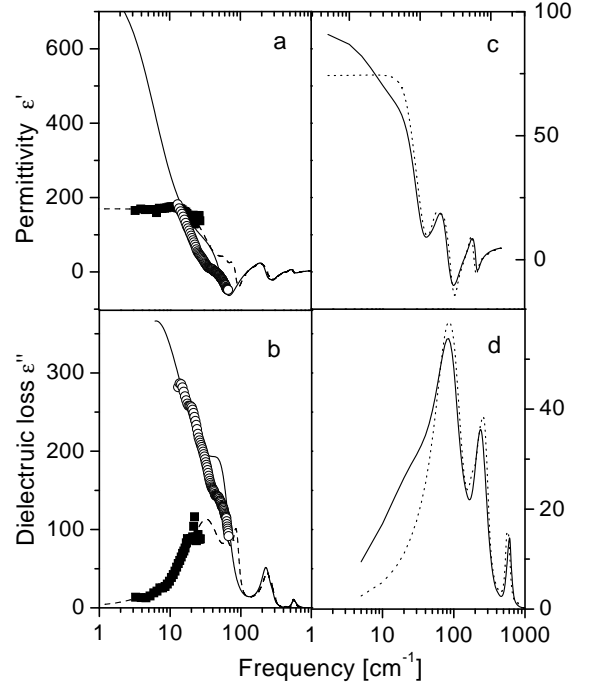}}
\caption{Experimental (a-b) and computed (c-d) dielectric
permittivity and losses: (a-b) squares denote experimental
microwave measurements obtained by using a THz spectrometer,
circles IR measurements; the dashed and solid lines are the
results of the fits to IR reflectivity at 20 K and 300 K
respectively; (c-d) the dashed and solid lines correspond to the
[001]$_{NCC'}$~ and [111]$_{NT}$~ structures respectively. }
\label{epsilon}
\end{figure}

\begin{figure}
\resizebox{0.45\textwidth}{!}{\includegraphics{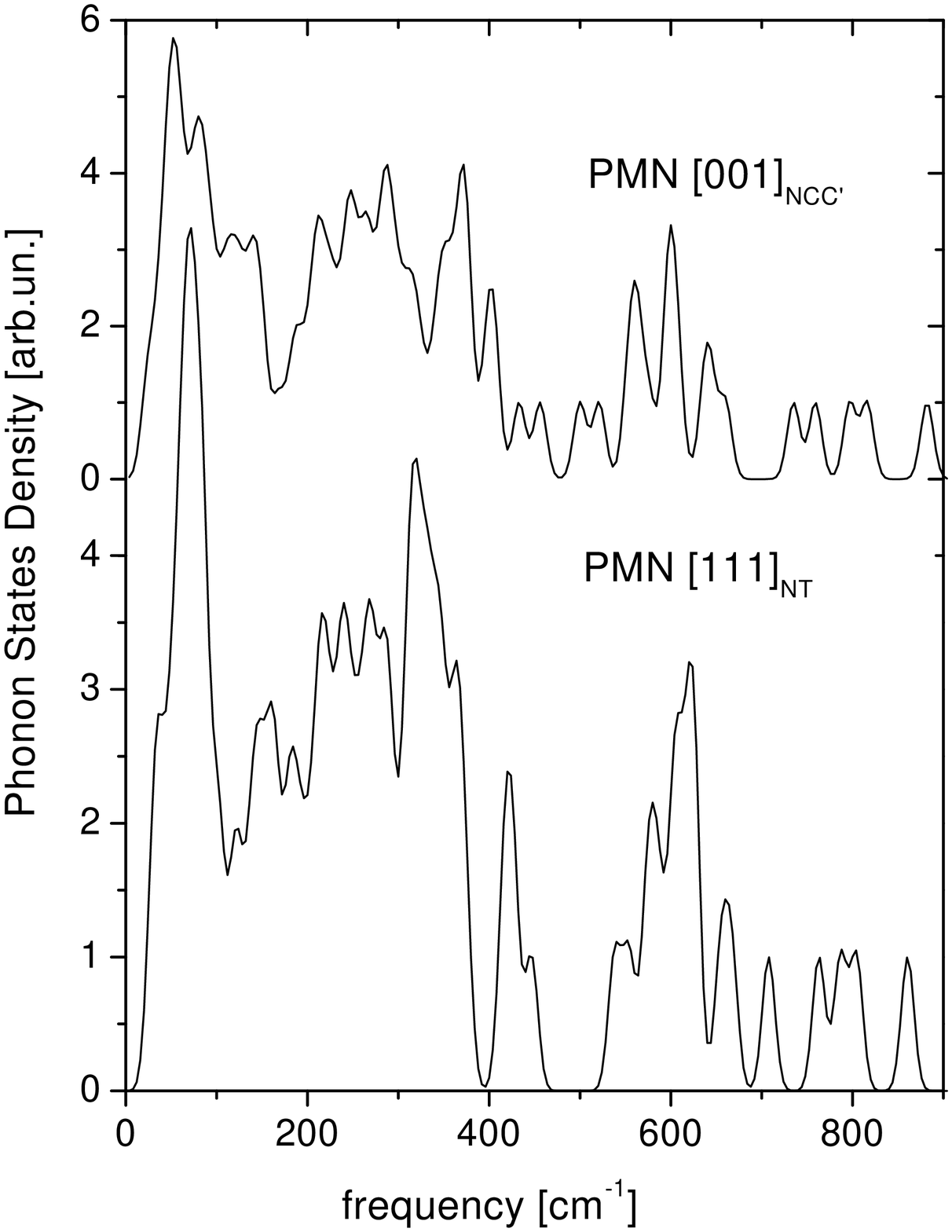}}
\caption{Computed zone-center phonon density of states in
structurally stable states.} \label{density}
\end{figure}

The IR reflectivity spectra are shown in Fig. \ref{reflectivity},
and compared with our experiment. It is seen that the best
agreement between the theory and experiment is for [111]$_{NT}$~
structure. The complex dielectric spectra in the phonon frequency
range are plotted in Fig. \ref{epsilon} together with experimental
data obtained. We have qualitative agreement for high frequencies.
At low frequencies, experiment shows very high values of
permittivity connected with relaxations of PNC. These frequencies
are below phonon's. Relaxation contributions to dielectric
permittivity were not included into our theoretical consideration
(we discussed these problems in a previous publication
\cite{KLT,KLT1}). Notice different scales in the experimental and
theoretical plots. The total zone-center phonon density of states
of the [111]$_{NCC'}$ and [111]$_{NT}$ structures are shown in
Fig.~\ref{density}. It is seen from this plot that the high energy
phonon band is not very dispersive compared to the low frequency
bands and compared to the bands known for ABO$_3$~ perovskites.
This can be explained by folding the bands due to ordering of Mg
and Nb.

Mode assignments were then made at several levels of approximation
by averaging the [111]$_{NT}$ DM.\cite{foot1} The lowest-frequency
modes of the supercell are related to the folded acoustic modes of
the parent 5-atom cell: excluding zero frequency zone center
acoustic modes, eigenvectors of the low-frequency modes in the
frequency interval 19-100 cm$^{-1}$~ are significant only on Pb.
This corresponds to extremely low-frequency acoustic antiphase Pb
displacements, which arise due to a very small diagonal element of
DM for Pb.

\begin{figure*}[!htbp]
\resizebox{1\textwidth}{!} {\includegraphics{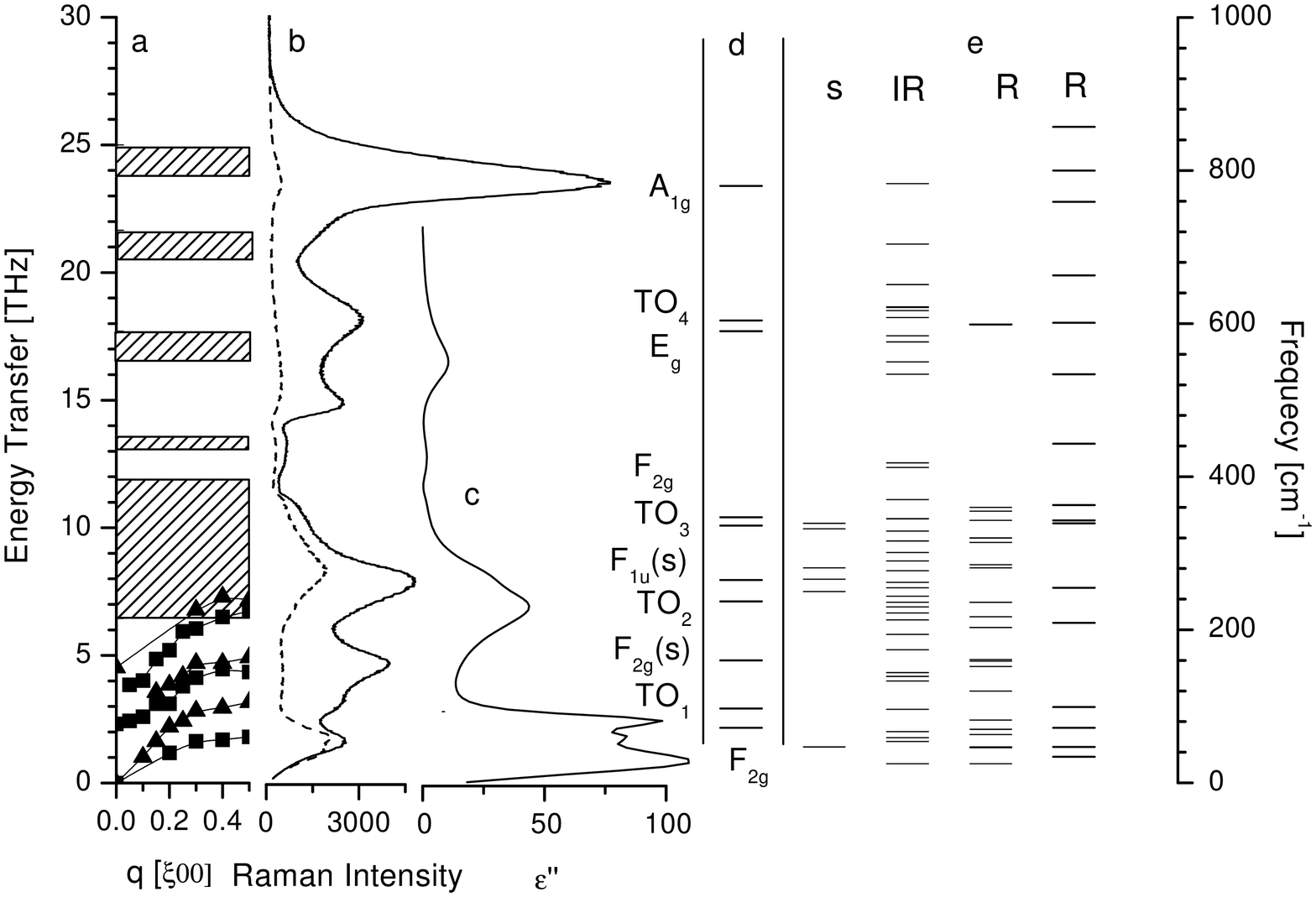}}
\caption{Comparison of low-temperature experimental data and
computational results: a) 12 K neutron scattering data
\cite{Dorner} (full triangles and squares correspond to the
longitudinal and transverse phonon dispersion branches; the bands
show the frequencies of continual neutron scattering); b) Raman
scattering data in two geometries (VV, solid line, and VH, dotted
line) recorder at 77 K; c) dielectric loss $\varepsilon''(\nu)$~
obtained from the fit of IR and THz spectra at 20 K;  d) Results
of first principles computations in $Fm\bar{3}m$~ symmetry; e)
Mode assignments for PMN [111]$_{NT}$~ structure in space group
Immm: two last columns correspond to different polarization of
Raman spectra. } \label{comparison}
\end{figure*}

Next, to understand how local 1:1 order may affect the phonon
properties, the [111]$_{NT}$~ dynamical matrix was averaged with
respect to a 1:1 ordered structure. The [111]$_{NT}$ structure is
the only one studied which is commensurate with such averaging.
Effectively, the [111]$_{Mg_{2/3}Nb_{1/3}}$ layers were treated as
though they contained only one average T-cation,
T=(Mg$_{2/3}$Nb$_{1/3}$). Symmetry analysis of [111]$_{NT}$~ in
$Fm\overline{3}m$, yields the following modes:
$A_{1g}(R)+E_g(R)+F_{1g}(s)+2F_{2g}(R)+4F_{1u}(IR)+F_{2u}(s)$.
Here, $IR$~ and $R$~ indicate IR- and Raman-active modes
respectively, and $s$~ indicates silent. Computed eigenvectors of
the [111]$_{NT}$ $P1$ structure were projected onto vectors of the
averaged computation to obtain the modes expected to have the
highest infrared and Raman activity. The results are shown in Fig.
\ref{comparison}, and compared with experimental data from
inelastic neutron scattering,\cite{Dorner} Raman scattering and
FIR reflectivity measurements (this study).

Polar TO modes [$F_{1u}(IR)$] are labeled TO1, TO2, TO3, and TO4.
The relative contributions from Pb, T, Nb, O$_{\parallel}$, and
O$_{\perp}$~ in the TO1 mode (100 cm$^{-1}$) are: -0.16, 0.21,
0.27, 0.32, and 0.41 \AA; where: O$_\parallel$~ is displaced along
B-O bonds; O$_\perp$~ is displaced perpendicular to B-O bonds.

The TO2 mode at 270 cm$^{-1}$, is predominately of the Slater
type.  Neutron \cite{Dorner} and our IR measurements (Table I)
exhibit bands near this frequency. The TO3 mode near 340
cm$^{-1}$~ is from Nb moving opposite to T. Experimental evidences
for this band is not definitive, although a mode is found near
this frequency in the fit to room-temperature IR data.  The TO4
mode near 580 cm$^{-1}$~ originates from O$_\parallel$-(T,Nb)
stretching, and it is clearly observed in the IR and neutron data.

In first-order Raman scattering from $Fm\overline{3}m$ crystals,
$A_{1g}$ and $E_g$ modes should be visible in VV geometry, while
$F_{2g}$ modes should be visible in VH geometry. The calculated
$A_{1g}$~ Raman active mode at 810 cm$^{-1}$~ matches quite well
the intense peak that is observed experimentally at 780 cm$^{-1}$
in VV spectra. In [111]$_{NT}$~ $Fm\overline{3}m$, this mode
corresponds to a fully symmetrical, IR inactive, O-breathing mode.
The calculated Raman active $E_g$ asymmetrical breathing mode near
590 cm$^{-1}$ is broadened due to local symmetry breaking, and
corresponds with experimentally observed bands at 500 and 600
cm$^{-1}$. In all cases, we had some frequency intervals (and even
a multipeak structures) when projecting the modes obtained for GSS
onto the modes derived from symmetrized DM.

The calculated Raman active $F_{2g}(2)$~ mode near 360 cm$^{-1}$~
involves mostly symmetrical O$_\perp$~ displacements. A wide
envelope of the bands is seen in our Raman experiment at this
frequency. The calculated low-frequency $F_{2g}$ Raman mode near
80 cm$^{-1}$ is mostly from Pb-displacements. In the VH spectrum,
the low-frequency band is a doublet (45 and 62 cm$^{-1}$). Both
components are close to the zone-boundary frequency of the TA
branch (50 cm$^{-1}$) according to neutron scattering data
\cite{Dorner} (Fig. \ref{comparison}).

Experimentally, there are strong Raman peaks near 150 cm$^{-1}$
and 250 cm$^{-1}$.  These peaks probably arise from local symmetry
breaking (octahedral tilting and/or polar distortions), and not
from Nb-T ordering.

\section{Discussion}\label{Discussion}

 All the PMN supercells studied here
(Fig. \ref{structures})
are unstable with respect to lower energy structurally stable states,
i.e $\Delta E_{Chem} > \Delta E_{SSS}$. Relaxation energies,
$\Delta E_{SSS}~-~\Delta E_{Chem}$, are large relative to the
thermal vibrational energy, which suggests that local
$displacive$~ symmetry breaking should persist to high
temperatures, generating local strain fields. Some instabilities,
however, have barriers of the order of thermal energy, especially
Last-type modes corresponding to Pb-BO$_6$~ stretching vibrations
in finite-size chemically ordered regions. The softest vibrations
are exhibited by Pb-ions that are coordinated by a highly
symmetric array of Mg- and Nb-ions, or Pb-ions surrounded by
Nb-ions only. Such Pb-ions have diagonal Pb-frequencies (the
ratio of the spring constant and mass) of only $\sim ~80~$
cm$^{-1}$.

Of the supercells studied here, [111]$_{NT}$, in $G_{SSS}~=~P1$,
has the lowest energy, and with respect to chemical SRO, it is the
best approximate of the RSM.  This suggests that it is also a good
approximate for chemically ordered domains in a PMN crystal with
$Pm\overline{3}m$~ global symmetry. Egami\cite{Egami1} and
Naberezhnov \textit{et al} \cite{Naberezhnov} reported that the
local ferroelectric instability in PMN below Burns temperature is
associated with Pb-displacements in opposition to the other ions
and this is essentially what occurs in the [111]$_{NT}$~ Last-mode
dominated FE instability, $Immm~\rightarrow~P1$. Note, however,
that Last-type instabilities occur in all five supercells.
Computed distributions of the interatomic distances in the 30-ion
supercells show that Pb-O distances split into two main groups at
2.5 \AA~and 3.2 \AA~, in agreement with experiment.\cite{Egami}

Computed dynamical charges in PMN are not as large as is typical
for simple ABO$_3$ perovskites\cite{Will}: the Nb charge varies
from 6.0 to 9.1. An FE-active ion that is surrounded by less
FE-active ions, e.g. Nb surrounded by Mg, typically has a reduced
dynamical charge, e.g. relative to Nb surrounded by Nb; and the
dynamical charge of the less FE-active ion increases (cf.
\cite{Wash00}). This happens because charge transfer is reduced
when an ion of relatively low electronegativity is surrounded by
ions with higher electronegativities. \cite{Osypenko}

Computed phonons for all the structures shown in
Fig. \ref{reflectivity} exhibit three similar features: (1)
B-O-B stretching modes at 500-900~cm$^{-1}$; (2) mixed B-O-B
bending and O-B-O stretching modes at 150-500~cm$^{-1}$; (3)
Pb-BO$_6$~ stretching modes at $\nu <$~150~cm$^{-1}$.
Differences between the computed IR spectra
for different supercells (Fig. \ref{reflectivity})
are purely quantitative, not qualitative. The [111]$_{NT}$~
spectrum is most similar to the experimental one: the
500-900~cm$^{-1}$~ band is relatively narrow, as in the experiment
(Fig. \ref{reflectivity}); but the other computed bands are less
intense than their experimental counterparts.  This is an artifact
of using the same damping constant for all computed modes (60
cm$^{-1}$~ is an average of the experimental values).  The 60
cm$^{-1}$~ damping constant, which we used in our computation, is
significantly larger than the usual value for pure perovskites,
$\sim$20 cm$^{-1}$. A large value for disordered PMN is expected
because the random fields caused by chemical disorder broaden the
phonon density of states. Experimentally, however (c.f. Sec.
\ref{IR}), some modes exhibited extremely small damping constants,
which implies sharper features in the reflectivity spectrum.

One can see in Fig. \ref{comparison} that the number and
distribution of the modes in $Immm$~ group over frequencies are
much richer than in $Fm3m$~ geometry. For example, the single
$A_{1g}$~ peak in $Fm3m$~ group can be connected with three $A_g$~
modes in $Immm$~ group (857, 800, and 759 cm$^{-1}$), in agreement
with Raman spectra. The distribution of the frequencies of these
peaks in $Immm$~ group is due to differences in frequencies for
oxygen ions in different Wyckoff positions: the stretching
vibrations of the oxygen ions in Nb-O-Nb- chains are harder than
other oxygen stretching vibrations, due to ferroelectric
instability in these chains (combined with nearest Pb ions) and a
large shift of oxygen ions in the equilibrium state (see Table
\ref{diag30}). There is also a difference between ionic vibrations
in Mg-O-Nb bridges and in Nb$_4$O$_4$~ squares in $xy$~ plane.

Ideally, $A_{g}$~ modes should be seen only in VV geometry. The
presence of some intensity in VH geometry can be regarded to local
fields and distortions. Notice the existence of an IR active mode
(704 cm$^{-1}$) in the same frequency interval which is due to a
large (minimal) size of the supercell in comparison with the unit
cell at $Fm3m$~ symmetry.

At the frequency of $E_g$~ mode of $Fm3m$~ group, there are
several Raman active symmetrical breathing $A_g$~ (663, 601, and
534 cm$^{-1}$) and one antisymmetrical breathing (599 cm$^{-1}$)
modes. The splitting and the absence of large intensity in VH
geometry in experimentally measured spectra in this frequency
interval testifies against $E_g$~ mode in $Fm3m$~ group but agrees
quite well with the computed splitting in $Immm$~ group and with
$A_g$~ symmetry of three of the four modes. In the same frequency
interval there is a wide distribution of IR active modes that well
corresponds to the wide peak in the FIR spectrum (TO4 mode in
terms of $Fm3m$~ symmetry).

The $Fm\overline{3}m$~ group assignment can hardly explain the
presence of a peak in Raman and IR spectra at a frequency a little
bit higher than 400 cm$^{-1}$. The computation within $Immm$~
symmetry does give both Raman active ($A_g$, 443, and $B_{2g}$,
423 cm$^{-1}$) modes as well as IR active modes (412, 418
cm$^{-1}$).

The position of $F_{2g}(2)$~ mode in $Fm3m$~ group is noticeably
higher than the position of the intensive peak in Raman spectrum
in VH geometry. The $Immm$~ group gives a few suitable modes in
the right place and gives a wide distribution of these modes over
frequencies that corresponds to a large width of the Raman lines
observed. However, at the present time, we cannot prove that the
peak of the Raman intensity in the computation corresponds to the
peak in the experiment, because we did not compute the
intensities.

Finally, we want to notice that there are a few low frequency
modes spread over the frequency interval 25-96~cm$^{-1}$ (below
the frequency of the soft mode) which are Raman and IR active.
These modes appear because of comparatively low symmetry of the
supercell. All they are in the range of the acoustic modes in the
reduced (5-ions) unit cell and are connected with antiphase Pb
displacements.

Taken together, the first principles results, and experimental
data from neutron scattering,\cite{Dorner} Raman, and IR
reflectivity enables assignments of specific vibrational modes to
experimental phonon peaks. Following Akbas and Davies \cite{Akbas}
PMN is approximated as the RSM. Positions of IR-active, Raman-active
and silent modes, were calculated, and distribution functions for
all these modes have finite width, and some are split.

There are a few low frequency modes in the range 25-96~cm$^{-1}$
(below the frequency of the soft mode) which are both Raman and IR
active. They reflect antiphase Pb-displacements \cite{Tkachuk}
that are allowed by the low supercell symmetry, and are all in the
acoustic-mode range for the reduced (5-ion) unit cell.

\section{Conclusions}

We studied five ordered supercells of PMN. Instabilities involving
Pb off-centering are observed in all supercells and instabilities
involving the tilting of O octahedra centered on Mg ions are
observed in most. When fully relaxed, the [111]$_{NT}$~ structure
described here has the lowest known first-principles energy for an
ordered PMN structure.  It is commensurate with the random site
model (RSM) of PMN\cite{Akbas}, and has short-range Mg-Nb
correlations that are most comparable to the RSM.  Its Raman and
IR spectra and pair distribution functions are in qualitative
agreement with experiment, allowing mode assignment of the
dominant features of the spectra. In its lowest-energy state, it
is polarized in approximately along an $[bbc]$ type direction,
yielding pseudomonoclinic symmetry. Unphysical aspects of this
perfectly ordered [111]$_{NT}$~ supercell as an approximant for
the 1:1 ordered regions in experimental PMN, include the lack of a
disordered matrix, the fact that there is no experimental evidence
for ordering on the Mg$_{2/3}$Nb$_{1/3}$ sublattice, and the
presence of infinite Nb-O-Nb... chains (which influence the
lattice dynamics).

\section*{Acknowledgments}
Y.I.Y. and R.S.K. appreciate support by the Grant NSF-DMR-0305588
and \#NSF-FY2004, S. K. and J. P. support by the Grant Agency of
Academy of Sciences (projects Nos. A1010203 and AVOZ1-010-914),
Grant Agency of the Czech Republic (projects No. 202/04/0993 ) and
Ministry of Education of the Czech Republic (project COST OC
525.20/00). S.A.P. grants ru.01.01.037 ("Russian Universities")
and 04-02-16103 (RFBR). S.B.V. acknowledges RFBR 02-02-16695 and
CRDF RP1-2361-ST-02.

\end{document}